\documentclass[aps,twocolumn,showpacs,amsmath,amssymb,prb,footinbib,floatfix,superscriptaddress]{revtex4}

\usepackage{graphicx}
\usepackage{dcolumn}
\usepackage{bm}

\usepackage{helvet}
\usepackage{amssymb}
\usepackage{amsmath}
\usepackage{amsfonts}
\usepackage{amssymb}
\usepackage{hyperref}
\usepackage{color}
\usepackage{verbatim}

\usepackage{xcolor}
\definecolor{red}{rgb}{0.7,0,0}
\definecolor{green}{rgb}{0.,0.35,0.}
\definecolor{blue}{rgb}{0.2,0.2,0.7} 
\definecolor{black}{rgb}{0.15,0.15,.15}

\makeindex

\begin{document}

\title{Critical properties and R\'enyi entropies of the spin-$3/2$ XXZ chain}

\date{\today}

\author{ M. Dalmonte}
\email{marcello.dalmonte@uibk.ac.at}
\affiliation{Institute for Quantum Optics and Quantum Information of the Austrian Academy of Sciences, A-6020 Innsbruck, Austria}

\affiliation{Dipartimento di Fisica dell'Universit\`a di Bologna and INFN, via Irnerio 46, 40126 Bologna, Italy}

\author{E. Ercolessi}
\affiliation{Dipartimento di Fisica dell'Universit\`a di Bologna and INFN, via Irnerio 46, 40126 Bologna, Italy}

\author{L. Taddia}
\email{luca.taddia2@gmail.com}
\affiliation{Dipartimento di Fisica dell'Universit\`a di Bologna and INFN, via Irnerio 46, 40126 Bologna, Italy}
\affiliation{IFT UAM/CSIC, 28049 Cantoblanco, Madrid, Spain}

\begin{abstract}
We discuss entanglement and critical properties of the spin-$3/2$ XXZ chain in its entire gapless region. Employing density-matrix renormalization group calculations combined with different methods based on level spectroscopy, correlation functions and entanglement entropies, we determine the sound velocity and the Luttinger parameter of the model as a function of the anisotropy parameter. Then, we focus on entanglement properties by systematically studying the behavior of R\'enyi entropies under both open and periodic boundary conditions, providing further evidence of recent findings about entanglement entropies of excited states in conformal field theory.
\end{abstract}

\pacs{75.10.Pq. 05.70.Jk, 05.10.Cc  }

\maketitle
 
\section{Introduction}

Since their first introduction\cite{heisenberg}, quantum spin chains have been an incredibly fertile field for theoretical physicists. They are in fact interesting for various reasons: first of all because of their one-dimensional (1D) nature, that enhances the importance of quantum fluctuations and forbids the application of mean-field or other ordinary perturbative approaches; secondarily because of  the integrability of some of them and the possibility of giving a descriptions of their low-energy sector by means of effective quantum field theories; lastly because of the availability of some numerical techniques that appear to be particularly powerful in these cases.

Spin-1/2 quantum spin chains with nearest neighbor interactions\cite{takahashi,bethe1931} have been widely considered in literature and, among them, the XXZ spin-1/2 chain is by far the most studied, both analytically and numerically. 
Being integrable via Bethe Ansatz \cite{bethe1931}, it represents a yardstick for non-exact techniques. Also, the physics of its low-energy sector is described in the continuum limit by a special class of Conformal Field Theories (CFTs)\cite{BPZ1984,difrancesco}  with conformal charge $c=1$, the so called Tomonaga-Luttinger Liquids (TLLs)\cite{tomonaga1950,luttinger1963,haldane1981} which represent a paradigm for all those models whose excitations are of bosonic nature\cite{bosonization, bosonization2}.

Higher spin XXZ chains are also very interesting, because they constitute examples of models for which, despite being non-integrable, one can provide quite well known established field theory descriptions\cite{schulz,affleckhaldane}. Moreover, their properties may be quantitatively determined with 
good accuracy by means of numerical simulations based on several
efficient methods such as exact diagonalization, quantum Monte Carlo, 
and Density-Matrix-Renormalization-Group (DMRG)\cite{bosonization2}. 
Thus, they represent a decisively indicative and efficient test in order to establish the validity of properties that, up to now, have been verified mainly in
integrable systems. In addition, the physics of these models can also be studied experimentally: for example, the spin-3/2 isotropic case is thought to model the behaviour of some kind of quasi-1D antiferromagnets of magnetic ions, such as CsVCl$_3$\cite{itoh1995} and AgCrP$_2$S$_6$\cite{mutka1993}, whereas various spin
models may be now engineered in cold matter setups of trapped ions\cite{schaetz2008,monroe2010,barreiro2011} and have promising future implementation 
with ultracold atoms and molecules in strongly anisotropic optical lattices\cite{bloch2008,micheli2006,gorshkov2011,dalmonte2011} and Rydberg atoms\cite{daley2010}.

In recent times, much attention has been devoted to the connection between Quantum Phase Transitions and entanglement properties in strongly correlated systems \cite{vidal2003,amico2008}. In particular, whenever the effective theory describing a system is conformal, it is well known that a fruitful way to get physical information from numerical simulations, especially from DMRG\cite{white,schollwock2005} calculations, is to look at quantities known as R\'enyi Entanglement Entropies (REs)\cite{laeuchli2008,nishimoto,xavier2010,xavier2011,dalmonteercolessitaddia}. More specifically, as it will be recalled in the following, for TLLs the knowledge of the REs yields a very careful estimation not only of the central charge of the underlying critical theory \cite{holzhey1994,calabrese2004,calabresecardy}, but also of the decay exponents of correlation functions\cite{calabresecampostrini,dalmonteercolessitaddia,xavieralcaraz}, which are encoded in the so called Luttinger parameter $K$.

The main aim of this work is to present a complete investigation
of critical and entanglement properties of the $S=3/2$ XXZ model all
over its gapless regime, and to compare the accuracy of different analysis methods 
employed to extrapolate thermodynamic quantities from finite-size 
numerical calculations based on the DMRG algorithm. 
In the first part, we will fully exploit the low-energy
field theory of XXZ model and determine its relevant quantities by considering different, independent observables; then, 
we will present a investigation of the entanglement 
entropies for bipartite intervals, considering both ground and excited states 
and systematically comparing the numerical findings with CFT 
predictions. The paper is structured as follows: in Sec. \ref{model}, we briefly review the main features of spin-$S$ XXZ chains from a field theoretical viewpoint,
focusing on the TLL universality class emerging in the half-integer $S$ case. We present our DMRG calculations and results on the $S=3/2$ model
in Sec.~\ref{results}, together with a brief resum\'e of all applied techniques.
In Section \ref{Renyis}, we perform a systematic investigation of the R\'enyi entropies of both the ground and excited states and compare the numerical 
findings with the predictions based on CFTs. Finally, we summarize the results and draw our conclusions in Sec.~\ref{conclusions}.

\section{Model Hamiltonian}\label{model}
The spin-$S$ anisotropic Heisenberg model, also known as XXZ chain, is described by the following
Hamiltonian\cite{bethe1931}:
\begin{equation}\label{H_XXZ}
H_{XXZ}=\sum_{i=1}^L(S_i^xS_{i+1}^x+S_i^yS_{i+1}^y+\Delta S_i^zS_{i+1}^z)
\end{equation}
where $\vec{S}_i$ is a spin-$S$ operator relative to the $i$-th site and $\Delta$ is the anisotropy coefficient. Here, $S$ can take positive half-odd-integer or integer values. In the simplest case $S=1/2$, the model is integrable by Bethe ansatz\cite{bethe1931,takahashi}, it is critical for $|\Delta|\leq 1$, and, in the interval $-1<\Delta\leq 1$ its low-energy physics  is effectively described by a conformal field theory with central charge $c=1$.

The picture becomes more puzzling as one moves away from the integrable $S=1/2$ case. At the isotropic point $\Delta=1$, 
chains with integer spin display a finite gap, whereas in the half-integer case the system is gapless and still described
by a $c=1$ CFT, as has been proved in a series of analytical and numerical studies~\cite{haldane,affleckhaldane,schulz,
alcarazmoreo,hallberg}. In such gapless regime, which persists in the finite range of interactions $-1<\Delta\leq1$, 
the low-energy spectrum is universally described by a TLL Hamiltonian\cite{tomonaga1950,
luttinger1963,haldane1981,bosonization,bosonization2}:
\begin{equation}\label{H_TLL}
 H=\frac{v_s}{2\pi}\int \; dx \left[(\partial_x \vartheta)^2/K+K(\partial_x\varphi)^2\right]
\end{equation}
where $\vartheta$ and $\varphi$ are conjugated density and phase bosonic fields, while the two interaction-dependent parameters
$v_s$ and $K$ are called sound velocity and Luttinger parameter respectively. Both the long-distance decay of correlation functions
and spectral properties are determined by $v_s$ and $K$: however, since no exact solution is known except for
$S=1/2$, one has to resort to unbiased numerical methods in order to quantitatively estimate the dependence of
such parameters with respect to $\Delta$. For the $S=3/2$ case, that we are going to extensively study in the following,
various numerical studies have been reported in literature, based on both exact diagonalization of small systems 
\cite{alcarazmoreo} and simulations based on the DMRG algorithm\cite{hallberg,xavier2010,xavieralcaraz}; 
we will systematically refer and compare our results to the known ones in the remainder of the paper.

\section{Numerical results: critical properties}\label{results}

In this section, we provide a complete study of the quantum critical regime $-1<\Delta\leq1$ for
the $S=3/2$ XXZ model employing the DMRG technique, which allows to estimate both 
ground state and excited state properties with notable accuracy. Being interested in various
physical quantities such as entropies, correlation functions and spectral properties, we employ
simulations with both open (OBC) and periodic (PBC) boundary conditions. While the former guarantee
better accuracy in the DMRG procedure, the latter are not affected by boundary effects; providing
estimates in both configuration represents a good check for our final results. In order to accurately
determine all relevant quantities of interest, we perform calculations with up to 512 (1156) states
per block for OBC (PBC), together with up to 5 sweeps for each intermediate size during the infinite-size
algorithm; in such a setting, typical discarded weights are of order $10^{-8}$ ($10^{-6}$)  during the 
last iteration.

Reliable estimates of the relevant physical quantities, $v_s$ and $K$, may be obtained with different
numerical analysis. In the following, we will employ three alternative methods based on independent
quantities: energy scaling in the low-energy spectrum, entanglement entropy and spin fluctuations.
A detailed account on how such quantities are related to the sound velocity and the Luttinger parameter
is given in each of the following subsections.

\subsection{Central charge}

\begin{figure}[t]{
\begin{center}
\includegraphics[width=0.75\linewidth]{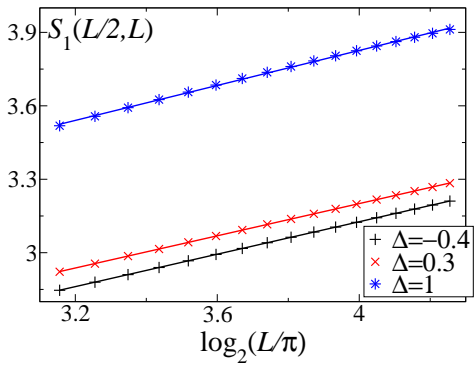}
\caption{ (Color online) Finite-size scaling of the bipartite von Neumann entropy as a function of the system size $L\in [28,60]$ for different
values of the anisotropy coefficient $\Delta$. Here, PBC are considered in order to discard oscillatory corrections.
Solid lines represent fits with Eq.~\ref{c_fit} yielding $c=0.995(4), 0.992(6)$ and $1.014(4)$ from top to bottom. }
 \label{fig:central_charge}
 \end{center}
 }
\end{figure}

As a first step in our study, we extract the central charge of the system from the scaling of
the block Von Neumann Entropy (VNE), which is defined as
\begin{equation}
 S_1 (l,L)=-\mbox{Tr}_A\rho_A\log_2\rho_A
 \end{equation}
Here, the system of total length $L$ is bipartite in two sub-systems $A,B$ of length $l,L-l$ respectively, and
$\rho_A$ denotes the reduced density matrix of $A$ with respect to $B$. In a CFT, one has~\cite{holzhey1994,calabrese2004,calabresecampostrini}:
\begin{equation}\label{S_cft}
S_{1}(l,L)=\frac{c}{3\eta}\log_2 [L\sin(\pi l/L)/\pi] + s_1 +S_1^{osc}
\end{equation}
where the coefficient $\eta=1,2$ for PBC/OBC, $s_1$ is a model-dependent constant and $S_1^{osc}$ represents finite-size oscillating corrections~\cite{calabresecampostrini}. In a finite system under PBC, the central charge can be determined by 
fitting the half-lattice entropy 
\begin{equation}\label{c_fit}
S_1(L/2,L)=\frac{c}{3} \log_2 (L/\pi) + s_1
\end{equation}
as a function of the system size $L$ by assuming a scaling form of the type $c(L)=c_0+a_0 L^{a_1}$. Typical results are plotted in Fig. \ref{fig:central_charge}: a best fit of
Eq. \ref{c_fit} for $L\in[28,60]$ gives an excellent agreement with the expected value, as $c_0=1.00$ up to a $2\%$ error over the entire parameter
range; for negative values of $\Delta$, an even better agreement is reached. Alternative techniques
to extract $c$ via finite-size scaling~\cite{xavier2010,nishimoto} lead to comparable results.
A good estimate of the central charge represents a reliable check of our numerical calculations,
and is also required to consistently perform a level spectroscopy analysis without targeting excitations in
momentum space~\cite{xavier2010,hallberg}.

\subsection{Sound velocity}
\begin{figure}[t]{
\begin{center}
\includegraphics[width=0.75\linewidth]{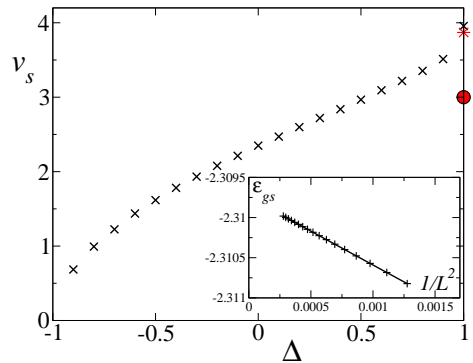}
\caption{(Color online) Sound velocity $v_s$ as a function of $\Delta$ as extracted from level spectroscopy methods (black crosses); at $\Delta=1$ the red 
circle denotes the spin wave results $v_s=3$, and a previous DMRG estimate~\cite{hallberg} $v_s=3.87$ is indicated by a 
red star. The maximum error is of order $6*10^{-2}$ at the antiferromagnetic point. Inset: typical finite size scaling of the ground state energy density $\epsilon_{gs}(L)$ for $\Delta=-0.5$. }
 \label{fig:sv}
 \end{center}
 }
\end{figure}

Following standard level spectroscopy methods based on the energy scaling of CFTs~\cite{difrancesco,xavier2010}, the sound velocity $v_s$ of 
the single component  TLL can be estimated by considering the finite-size scaling of the ground state energy density, which under PBC
reads as:
\begin{equation}
\epsilon_{gs}(L)=\epsilon_0+\frac{v_s c\pi}{6L^2}+...
\end{equation}
$\epsilon_0$ being  the energy density in the thermodynamic limit, and employing the previously found values of the central charge.
The large number of system sizes considered in our calculations allows to safely perform a four-parameter fit of the form
$\epsilon_{gs}(L)=a_0+a_1/L^2+a_2/L^{a_3}$; typical fitting results are shown in the inset of Fig.~\ref{fig:sv}. At the
antiferromagnetic point $\Delta=1$, our results are in good agreement with a previous DMRG study\cite{hallberg}, where $v_s$ was 
extracted by targeting the first excited states at finite momentum, and in sharp disagreement with the spin wave result\cite{affleck1987},
$v_s^{SW}=3$, in which quantum fluctuations are only approximately treated. Including logarithmic corrections
according to Wess-Zumino-Novikov-Witten\cite{bosonization,hallberg} theory does not lead to appreciable differences.

The complete dependence of $v_s$ versus $\Delta$ is plotted in Fig.~\ref{fig:sv} and suggests how, once approaching
the ferromagnetic phase transition, the velocity of the sound excitations seems to approach zero. On the other hand,
$v_s$ reaches its maximum value at the Berezinskii-Kosterlitz-Thouless point\cite{bosonization}, in full analogy with the $S=1/2$ case\cite{takahashi,bosonization}.

\subsection{Luttinger parameter}

When dealing with models whose low-energy physics is captured by the TLL Hamiltonian, 
the parameter $K$ represents a fundamental quantity as it determines the long-distance
decay of all correlation functions and thus which susceptibilities are the most relevant  in the
microscopic model\cite{bosonization2}. We will thus employ three complementary methods to extract such a quantity
from numerical simulations. This procedure allows to systematically check the validity of
each method in various parameter regimes, and, close to the antiferromagnetic point, 
indicates which type of estimate is more affected by the well known logarithmic corrections.

\begin{figure}[t]{
\begin{center}
\includegraphics[width=0.75\linewidth]{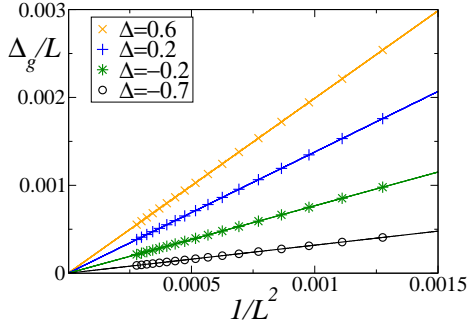}
\caption{(Color online) Finite size scaling of the energy gap $\Delta_g$ as a function of $1/L^2$ for different
values of $\Delta$; solid lines are best fits (see text). The Luttinger parameter estimated via Eq.
~\ref{delta_g} is, from top to bottom, $K=2.44(0),2.95(9),3.69(4),6.03(1)$.}
 \label{K_LS}
 \end{center}
 }
\end{figure}

\begin{figure}[t]{
\begin{center}
\includegraphics[width=0.85\linewidth]{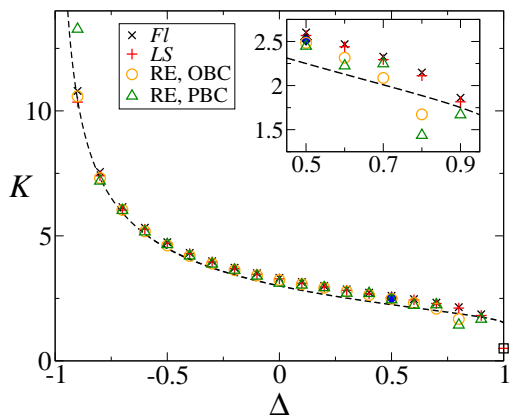}
\caption{(Color online) Estimate of the Luttinger parameter as a function of the anisotropy parameter. 
In the legend, $Fl, LS, RE$ denote spin fluctuations, level spectroscopy and RE results with OBC or PBC.
The dashed line is the conjecture of Ref. \onlinecite{alcarazmoreo}. The
blue diamond ($\Delta=0.5$) and the black square ($\Delta=1$) indicate the LS and exact results given in  Ref.~\onlinecite{xavier2010}
and \onlinecite{schulz} respectively.  The inset shows a magnification of the region close to the antiferromagnetic 
point: here, REs do not provide fully reliable results due to the presence of additional finite-size corrections,
while predictions from LS and spin-fluctuations are in excellent agreement. In both panels, numerical errors are 
smaller than the size of the symbols, except for the $\Delta=-0.9$ estimate based on RE with PBC: in this
case, the absolute numerical error due to the fitting procedure is of order 2 due to a very strong size-dependence of the fits. }
 \label{K}
 \end{center}
 }
\end{figure}

As a first way to estimate $K$, we make use of the previously calculated $v_s$ and apply
the level spectroscopy method by targeting the first excited state with total magnetization 
$\langle \sum_iS^z_i\rangle=1$ under PBC. Given its energy density $\epsilon_{+1}(L)$,
the energy gap $\Delta_g(L)=L(\epsilon_{GS}(L)-\epsilon_{+1}(L))$ scales to zero in the 
thermodynamic limit as~\cite{bosonization2,xavier2010}:
\begin{equation}\label{delta_g}
\Delta_g(L)= \frac{\pi v_s }{2 K L} +...
\end{equation}
We thus performed a least-square regression fit as a function of $1/L^2$ and a non-linear
fit of the form $\Delta_g(L)=b_0/L^2+b_1/L^{b_2}$ in order to estimate the effect of higher-order
corrections, which turns out to be negligible if $|\Delta|\leq0.9$, as can be inferred from
typical finite-size scalings  shown in Fig.~\ref{K_LS}. At the antiferromagnetic point, however, 
the quality of the best fit with just algebraic contributions turns out to be insufficient due 
to the presence of strong logarithmic corrections\cite{igloi2006}. We thus apply the same fitting procedure 
of Ref.~\onlinecite{hallberg},  which takes into account  logarithmic corrections as ensuing from the 
underlying SU(2) Wess-Zumino-Novikov-Witten field theoretical structure\cite{bosonization,tsvelik}, obtaining
$K=0.499\pm0.005$ at the critical point, in good agreement with previous numerical\cite{hallberg} and
analytical findings\cite{schulz}. A summary of the so obtained  results all over the critical regime is 
presented in Fig.~\ref{K} and will be discussed at the end of this subsection.

\begin{figure}[t]{
\begin{center}
\includegraphics[width=0.75\linewidth]{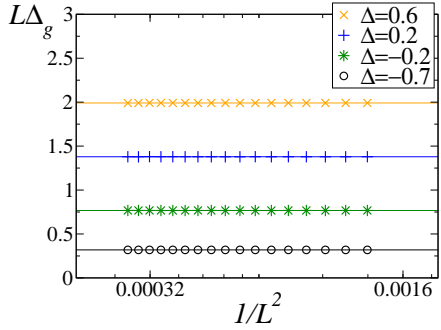}
\caption{(Color online) Finite size scaling of the rescaled energy gap $\Delta_g$ as a function of $1/L^2$ (in
logarithmic scale) for the same anisotropies considered in Fig.~\ref{K_LS}: solid lines are best fits (see text), showing how additional
finite-size contributions are usually negligible.}
 \label{K_LS2}
 \end{center}
 }
\end{figure}

\begin{figure}[t]{
\begin{center}
\includegraphics[width=0.75\linewidth]{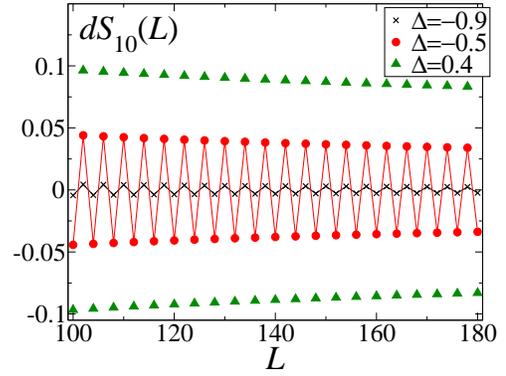}
\caption{(Color online) Oscillating factor of the $\alpha=10$ RE as a function of the system size $L$ for OBC. The amplitude
of the oscillations strongly decreases when approaching the ferromagnetic point\cite{xavieralcaraz}: even though, an accurate estimate
of $K$ is still possible (with typical relative error around $10^{-2}$) by considering the appropriate entanglement entropy. }
 \label{K_RE}
 \end{center}
 }
\end{figure}

\begin{figure}[b]{
\begin{center}\includegraphics[width=0.75\linewidth]{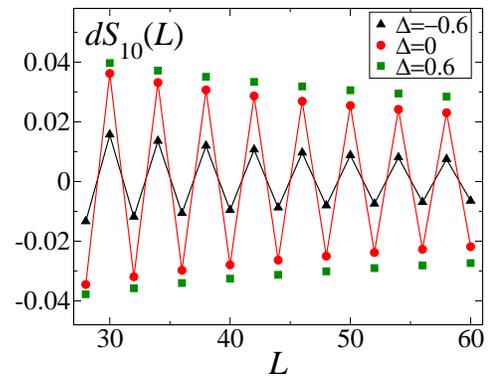}
\caption{(Color online) Oscillating factor of the $\alpha=10$ RE as a function of the system size $L$ for PBC. 
Continuous lines are guides for the eye.
 }
 \label{K_RE_PBC}
 \end{center}
 }
\end{figure}

The second way of getting $K$ is the one introduced by the authors in Ref. \onlinecite{dalmonteercolessitaddia}, based
on the finite size corrections of the bipartite R\'enyi entropies:
\begin{equation}\label{REE}
 S_\alpha(l,L)=\frac{1}{1-\alpha}\log_2\mbox{Tr}_A\rho_A^\alpha
\end{equation}
whose scaling, for a CFT with $c=1$, follows\cite{calabresecampostrini}:
\begin{equation}\label{cc}
S_\alpha(l,L)=\frac{c(1+\frac{1}{\alpha})}{6\eta} \log_2 [L\sin(\pi l/L)/\pi] + s_{\alpha}+S^{osc}_{\alpha}(l,L)
\end{equation}
where for $\alpha\rightarrow 1$ one recovers the von Neumann entropy (\ref{S_cft}). The influence of the Luttinger parameter
on such entanglement entropies is encoded in the oscillating factor, which on a finite chain of size $L$ scales as
\cite{calabresecampostrini,xavier2010,dalmonteercolessitaddia}:
\begin{equation}\label{FSC}
 S_\alpha^{osc}(l,L)=\cos(2k_Fl+\omega)\frac{f_\alpha\left(\frac{l}{L}\right)}{\left|2\sin k_F\frac{\eta L}{\pi}\sin\frac{\pi l}{L}\right|^\frac{2K(L)}{\eta\alpha}}
\end{equation}
where $k_F=\pi/2$ is the Fermi momentum and $f_{\alpha}(l/L)$ is a scaling 
function~\cite{calabresecampostrini}. In order to circumvent finite size corrections to the
central charge and non-universal features related to $f_{\alpha}$, one can 
consider the following entropy difference:
\begin{equation}\label{ourformula}
 dS_\alpha(L)=S_\alpha\left(\frac{L}{2},L\right)-S_\alpha\left(\frac{L}{2}-\frac{\pi}{2k_F},L\right)
\end{equation}
which, for $L\gg 1$, reduces to\cite{dalmonteercolessitaddia}:
\begin{equation}
 dS_\alpha(L)=\frac{\pi^4\left(1+\frac{1}{\alpha}\right)}{48\eta\ln 2k_F^2}\frac{1}{L^2}+\frac{\cos\left(k_F L\right)}{L^\frac{2K}{\eta\alpha}}\left(a+O\left(\frac{1}{L}\right)\right).
\end{equation}
The finite size scaling of $dS_{\alpha}$ with respect to the system size $L$ allows to precisely estimates
$K$ under both OBC and PBC, as discussed in detail in Ref.~\onlinecite{dalmonteercolessitaddia}\footnote{Notice that 
in the limiting cases were no oscillations are present, such method cannot be trivially applied\cite{xavier2011}. However, in the context of spin chains in Luttinger liquid regimes, a finite $k_F$ usually ensures the applicability of such criterion\cite{dalmonteercolessitaddia,xavieralcaraz}.}. 
Since it has been noticed that the oscillation amplitude is very small for Heisenberg chains with $S>1/2$
and $\alpha\simeq 2$, we employed REs with larger values of $\alpha$ in order to get a reliable estimate
from the DMRG results. In particular, our estimate of $K$ are mostly based on the $\alpha=10$ RE,
as these data represent a good compromise between small oscillation amplitude and slow
oscillation decay\footnote{A large value of $\alpha$ requires larger system sizes to get a reliable 
estimate of the oscillation decay exponent.}. 

Typical results from both PBC and OBC with systems sizes $L\in[28,60] (L\in[100,180])$ are presented
in Figs.~\ref{K_RE} and \ref{K_RE_PBC}. We notice that even though it  considerably decreases close to the ferromagnetic
transition, the magnitude of the oscillation is still large enough to perform accurate finite-size scaling. For $\Delta>0$, comparable results
may be obtained even with smaller $\alpha$'s. However, close to the antiferromagnetic point, different
types of corrections arise and the quality of the fitting procedure rapidly decreases. Thus, extracting
$K$ beyond $\Delta=0.7$ turns out to be numerically challenging. As can be seen from Fig.~\ref{K}, data
points obtained via $dS_{\alpha}$ significantly deviates (of about a $5-15\%$ discrepancy) from other estimates 
in this regime. Finally, let us point out that, for $\Delta=-0.9$, the decay exponent for PBC is so large that oscillations
are strongly dumped for large system sizes, so that an accurate estimate of $K$ is not possible; for OBC instead,
being the exponent smaller by a factor of $2$, the estimate is still in very good agreement with all other methods
even for $\Delta=-0.9$.

\begin{figure}[t]{
\begin{center}
\includegraphics[width=0.8\linewidth]{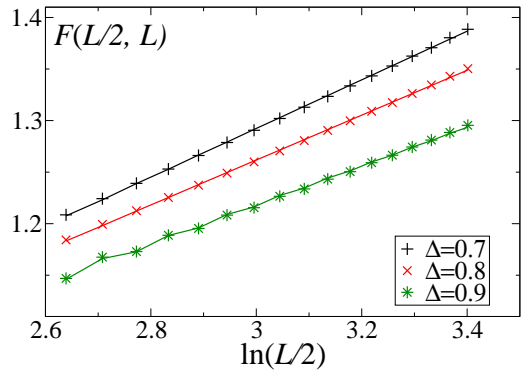}
\caption{(Color online) Spin fluctuations as a function of the system size $L$ under PBC. Solid lines are
best fits with Eq.~\ref{fl_fit}.}
 \label{K_fl}
 \end{center}
 }
\end{figure}

The third way we estimate $K$ is through spin fluctuations\cite{songrachellehur}. For a bipartition as the one used to define the von Neumann entropy, we define the quantity $F=\left<\left(\sum_{i\in A}S^z_i\right)^2\right>-\left<\sum_{i\in A}S^z_i\right>^2$. For a TLL under PBC, spin fluctuations behave as\cite{bosonization2,songrachellehur,dalmonteercolessitaddia}:
\begin{equation}\label{fl_fit}
 F(l, L)=\frac{K}{\pi^2}\ln\left[ \frac{L}{\pi}\sin\left(\frac{\pi l}{L}\right)\right]+A_1+\mathcal{O}\left(l^{2K}\right)
\end{equation}
and provide a quantitative estimation of $K$\cite{dalmonteercolessitaddia,songrachellehur,rachellaflorenciesonglehur}. 
We employ such method by fitting the half-lattice spin fluctuation $F(L/2,L)$ as a function of the system size
in the interval $L\in[28,60]$; typical fitting results are illustrated in Fig.~\ref{K_fl}. At the antiferromagnetic
point $\Delta=1$, logarithmic corrections to correlation functions severely modify Eq.~\ref{fl_fit}: strong oscillations,
not captured by a quadratic TLL theory, emerge with respect to the parity of $L$, preventing a reliable estimate
of $K$.

The results obtained in this section are all shown in Fig.~\ref{K},
from which one can see that they agree over almost the whole anisotropy parameter range.
In general, the level spectroscopy method leads to more accurate estimates, as the quantities
it relies on are very accurately estimated with DMRG, and additional finite size corrections 
seems to play a minor role; on the contrary, REs estimates, which are extremely precise for $\Delta<0.6$,
may indeed suffer from both larger truncation errors and stronger finite-size dependences close
to the antiferromagnetic point, resulting in a worst mutual agreement with fluctuations and LS.
Finally, all results are compared with a previously 
stated conjecture based on exact calculations on small system sizes\cite{alcarazmoreo} which relates the Luttinger parameter $K_S$ of a spin-$S$ 
Heisenberg chain with $S=1/2$ case in the $\Delta<0$ regime:
\begin{equation}\label{amconj}
 K(\Delta)_S=2SK(\Delta)_{1/2}.
\end{equation}
Remarkably, as it can be seen from Fig.~\ref{K}, this conjecture appears to be in a semi-quantitative agreement with the numerical results even well beyond its original validity regime. Discrepancies among the results emerge mainly  close to the antiferromagnetic point, where logarithmic corrections differently affect the techniques we have presented: in this regime, level spectroscopy 
turns out to be the most reliable method to extract the Luttinger parameter, as field theoretical instruments
allow to perform more accurate scaling hypothesis with respect to methods based on REs and fluctuations. Nevertheless, the picture suggests that a deeper theoretical insight on the analytical properties of  both REs and fluctuations close to critical point with logarithmic corrections may in principle enlarge their regime of applicability.

\section{Numerical results: Entanglement entropies and connection with CFTs}\label{Renyis}

\subsection{R\'enyi entropies of excited states}\label{excited states}

\begin{figure}[t]{
\begin{center}
\includegraphics[width=7.23cm]{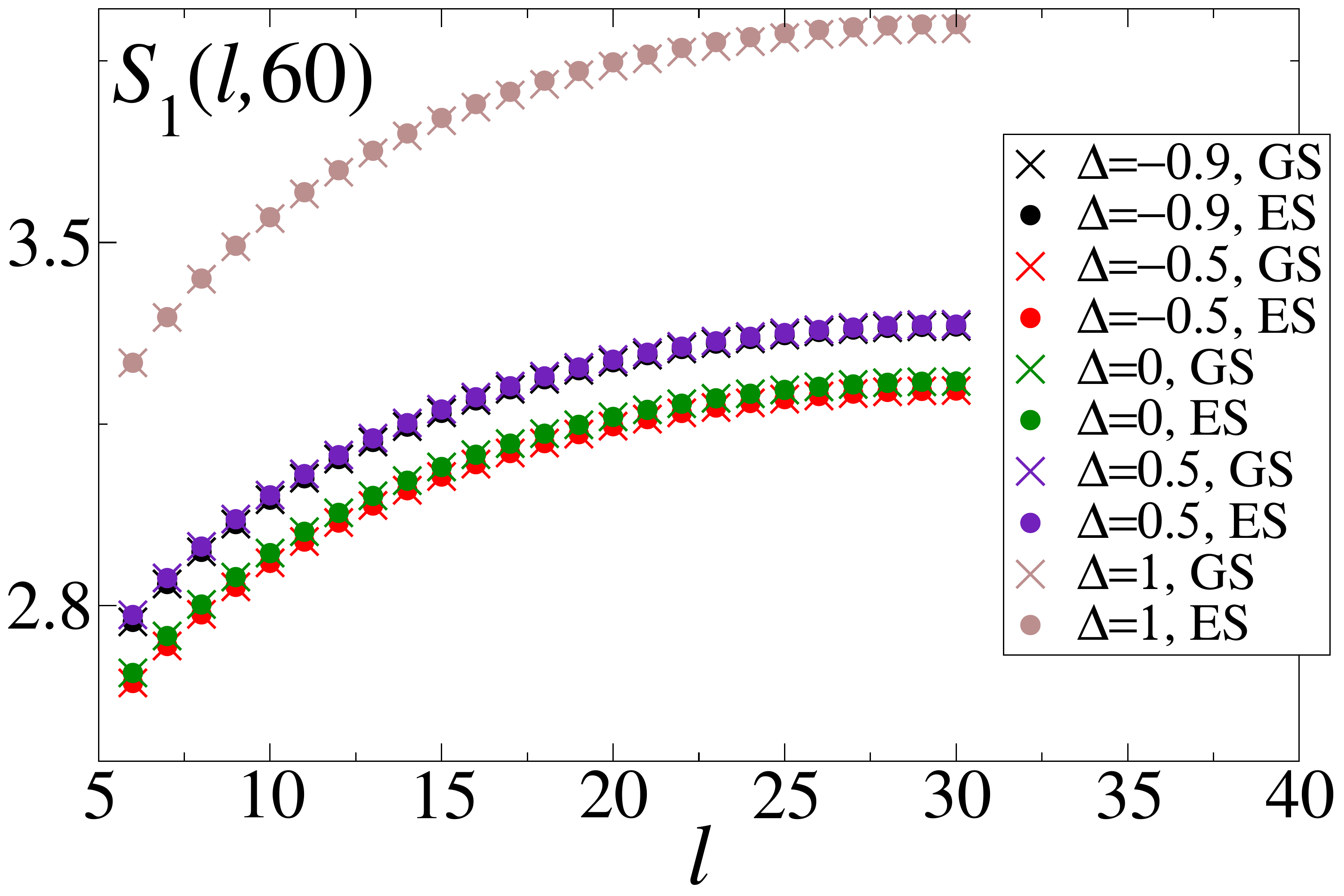}
\includegraphics[width=7.23cm]{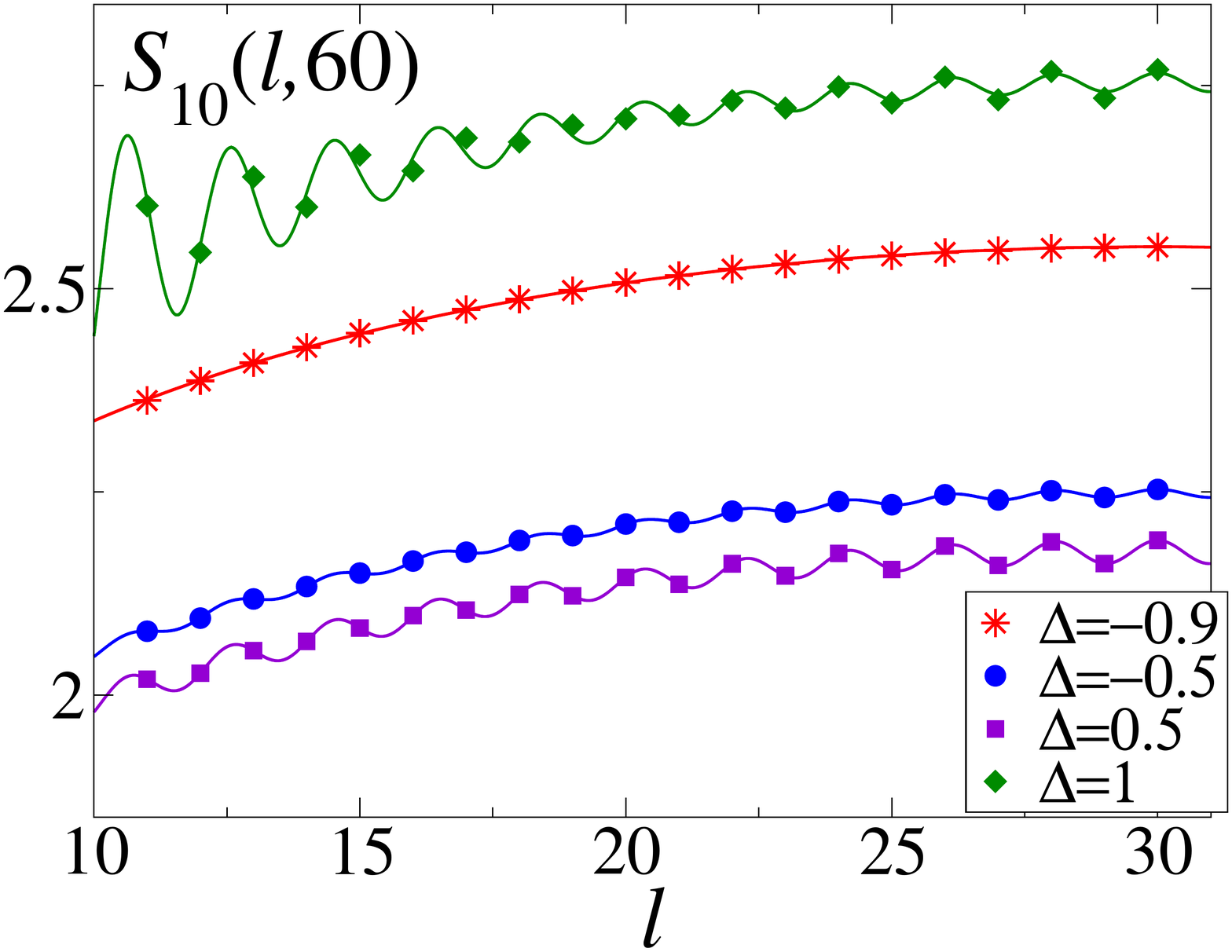}
\caption{(Color online) Upper panel: von Neumann entropies of ground and excited states 
for several values of $\Delta$ as a function of the block length $l$; here, $L=60$ and PBC 
are considered. Lower panel: $\alpha=10$ R\'enyi entropies of excited states at $L=60$. Solid lines are best fits obtained using Eq.~\ref{FSC} with $k_F=31\pi/60$.}
 \label{VNEEs_gs-exc}
 \end{center}
 }
\end{figure}

In recent times, an analytical formula for the REs of low-energy excited states in a CFT was derived and numerically verified for various quantum spin chains\cite{alcarazberganzasierra,alcarazberganzasierra2}. In particular, it has been predicted that the trace of the reduced density matrix $\rho_{A,\Upsilon}^\alpha$ of excited states generated by primary operators $\Upsilon$ of conformal weights $h,\ \bar{h}$  
satisfies the following relation:
\begin{eqnarray}\label{exc}
 && \alpha^{2\alpha(h+\bar{h})}\mbox{Tr}_A\rho_{A,\Upsilon}^\alpha=\\
&=&\frac{Z(\alpha)}{Z(1)^\alpha}\frac{\left<\prod_{j=0}^{\alpha-1}\Upsilon(2\pi j/\alpha)\Upsilon^\dagger(2\pi(j+l)/\alpha)\right>_{\mbox{cy}}}{\left<\Upsilon(0)\Upsilon^\dagger(2\pi l)\right>^\alpha_{\mbox{cy}}}.\nonumber
\end{eqnarray}
Here, $\alpha$ is assumed to be a positive integer, but the result can be analytically 
continued all $\alpha\geq1$; $Z(\alpha)$ is the partition function on a torus of
dimensions $2\pi\alpha$ and $2\log_2[L/\pi\sin(\pi l/L)]$; $\left<\cdots\right>_{\mbox{cy}}$
denotes expectation value on the vacuum state on a cylinder of length $2\pi$. 
In particular, if $\Upsilon$ is a vertex operator, formula (\ref{exc}) predicts that 
the VNE (and, up to oscillating terms, all REs)  of the excited state generated by $\Upsilon$ should be equal to the one
of the ground state.
This finding has been numerically verified in a series of exactly solvable
 spin models in Ref.\onlinecite{alcarazberganzasierra,alcarazberganzasierra2}. In order to further strengthen it, we considered the
REs of the state generated by applying a vertex operator on the ground state (which belongs to the $S^z_{tot}=\sum_{j=1}^LS_j^z=0$ sector)
thus obtaining the ground state in the  $S^z=1$ sector.
Spanning the entire critical region and employing PBC, we found excellent
agreement with the CFT prediction for both von Neumann and $\alpha>1$ REs.
In the former case, the entropy of excited and ground state coincide within numerical
uncertainty up to a constant shift of order $10^{-2}$, as can be seen from typical data 
presented in the upper panel of Fig.~\ref{VNEEs_gs-exc}; for small systems up to
$L=12$, we further checked this behavior with exact diagonalization.

We afterwards checked the relation between the oscillation corrections and the 
Luttinger parameter as extracted in the previous section by considering REs with $\alpha>1$.
Even in this case, the agreement with the predicted behavior is remarkable,
except at the ferromagnetic point where the quality of the fit significantly decreases. 
Results of $S_{\alpha}^{ex}(l, 60)$ as a function of $l$ are plotted in Fig.~\ref{VNEEs_gs-exc},
lower panel, and suggest that the amplitude of the oscillations follows a similar
behavior as in the ground state case, namely oscillations are more pronounced
when approaching the antiferromagnetic point.

\subsection{Entanglement behavior of Heisenberg chains with different $S$}\label{xavier_sec}

In a recent paper, Ref.\onlinecite{xavier2010}, an interesting property concerning entanglement entropies of XXZ chains was conjectured. Based on numerical DMRG simulations, it was shown that, under both OBC and PBC, the von Neumann entropy of XXZ spin chains with different half-odd-integer $S\geq 5/2$ satisfies the following relation:
\begin{equation}\label{xc_eq}
(\ln 2) \Delta S_1(S)\equiv S_1(l,L)_S-S_1(l,L)_{S-1}=\frac{1}{2S-1}+\epsilon_S
\end{equation}
where $\epsilon_S\rightarrow 0$ in the $L\rightarrow\infty$ limit. Such relation implies that, independently
on the anisotropy $\Delta$, the non-universal constant contribution acquires a universal form;
interestingly, it scales as the inverse of the difference between the one-site Hilbert space dimension
of the $S$ case minus the $S=1/2$ one. 

\begin{figure}[t]{
\begin{center}
\includegraphics[width=0.75\linewidth]{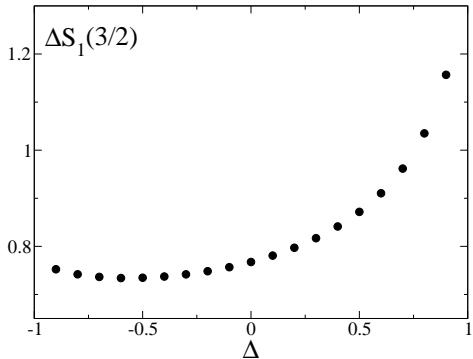}
\caption{$\Delta$-dependence of $\Delta S_1(3/2)$.}
 \label{XC}
 \end{center}
 }
\end{figure}

We systematically investigated the behavior of $\Delta S_1(3/2)$ as a function of $\Delta$ 
by employing PBC in order to get rid of the oscillation contribution, which, as noticed in 
literature\cite{xavieralcaraz} and confirmed in the previous section, is extremely different
for different values of $S$. As a preliminary check, we compared our value of $\Delta S_1(3/2)$
at $\Delta=1/2$ with the one reported in Ref.\onlinecite{xavier2010}, finding indeed 
very good agreement up to the different logarithmic basis employed here. 
Due to a very slight $l$-dependence of $\Delta S_1(3/2)$, we mostly considered its
mean value in the interval of block length $l\in[11,29]$ in systems with up to $L=60$
sites. A schematic plot of the entropy difference at $L=58$ is presented in Fig.~\ref{XC}: $\Delta S_1 (3/2)$
displays a notable non-monotonic $\Delta$-dependence.
Similar plots are obtained with different $L$.
In summary, our results firmly confirm that Eq.~\ref{xc_eq} does not hold for $S=3/2$, in accordance
with Ref.~\onlinecite{xavier2010}.

\section{Conclusions}\label{conclusions}
In this work, we analized critical and entanglement properties of the spin-$3/2$ anisotropic
XXZ chain. By means of DMRG simulations, we systematically calculated the sound 
velocity and the Luttinger parameter in the entire critical region with a variety
of methods such as level spectroscopy, entropy analysis and spin-fluctuations. At the 
antiferromagnetic point, logarithmic corrections prevent an accurate estimate of the Luttinger
parameter via spin-fluctuations and entropy oscillations, whilst level spectroscopy, where
such corrections may be systematically introduced, still provide reliable results. Such findings
benchmark the use of level spectroscopy techniques, which stem as preferable over correlation function methods
based on fluctuations and R\'enyi entropies when approaching and determining phase transition points with logarithmic
corrections. Away from such delicate regimes, all methods give compatible results,
in agreement with previous studies, although REs usually require more accurate calculations
since their absolute value is very small in our case study.

Finally, we investigated in detail the behavior of R\'enyi entropies of both ground and 
excited states. In the former case, we compared the REs of the $S=1/2$ and $S=3/2$
cases, proving that they are not connected by general relations as in the $S>3/2$ case.
Then, we provided evidence that recent results on the RE of certain excited states in
a conformal field theory are verified in the model of interest, presenting the first numerical
evidence of such expected behavior in non-integrable models. Even though the leading 
corrections to REs of excited states and those of the ground state display the same type of behaviour\cite{xavier2011,alcarazberganzasierra2}, whether the combination  of  the two may allow for a precise determination of  physical quantities of interest still stands as an open question.\\

\section*{Acknowledgments}

We thank F. C. Alcaraz, S. Evangelisti, M. Ibanez, F. Ravanini and G. Sierra for fruitful discussions, F. Ortolani for help with the DMRG code, and G. Ramirez for technical support. This work was supported in part by the INFN COM4 grant NA41. M. D. acknowledges support by the European Commission via the integrated project AQUTE.

 \end{document}